\documentclass[line numbers]{aa}
\usepackage{graphicx}
\usepackage{txfonts,twoopt}

\usepackage{hyperref}

\hypersetup{
    colorlinks=true,
    linkcolor=blue,
    citecolor = blue,
    filecolor=magenta,
    urlcolor=blue}
\usepackage{booktabs}
\usepackage[switch]{lineno}
\begin{document}

   \title{The star formation history of the first bulge fossil fragment candidate Terzan 5}

   \author{C.~Crociati
          \inst{1,2,3}
          \and
          M.~Cignoni\inst{3,4,5}
          \and
          E.~Dalessandro\inst{3}
          \and
          C.~Pallanca\inst{2,3}
          \and
          D.~Massari\inst{3}
          \and
          F.~R.~Ferraro\inst{2,3}
          \and
          B.~Lanzoni\inst{2,3}
          \and
          L.~Origlia\inst{3}
          \and
          E.~Valenti\inst{6,7}          
          }

   \institute{Institute for Astronomy, University of Edinburgh, Royal Observatory, Blackford Hill, Edinburgh EH9 3HJ, UK\\
              \email{ccrociat@ed.ac.uk}
         \and
             Dipartimento di Fisica e Astronomia, Universit\`a di Bologna, Via Gobetti 93/2 I-40129 Bologna, Italy
         \and
             INAF-Osservatorio di Astrofisica e Scienze dello Spazio di Bologna, Via Gobetti 93/3 I-40129 Bologna, Italy
         \and
            Dipartimento di Fisica, Universit\`a di Pisa, Largo Pontecorvo, 3, I-56127 Pisa, Italy
         \and
            INFN, Largo Pontecorvo 3, I-56127 Pisa, Italy
         \and
            European Southern Observatory, Karl-Schwarzschild-Strasse 2, 85748 Garching bei M\"{u}nchen, Germany
         \and
            Excellence Cluster ORIGINS, Boltzmann-Strasse 2, D-85748 Garching bei M\"{u}nchen, Germany
             }

   \date{Received June 20, 2024; accepted October 18, 2024}

 
  \abstract
   {Terzan~5 and Liller~1 are the only bulge stellar clusters hosting multi-iron and multi-age stellar populations. They are therefore claimed to constitute a novel class of astrophysical objects: the fossils of massive star-forming clumps that possibly sank to the center of the Milky Way and contributed to the formation of the bulge. This is based on the hypothesis that the ancient clumps were able to retain iron-enriched supernova ejecta, later giving rise to younger and more metal-rich populations. }
   {A way to investigate this scenario is reconstructing their star formation histories (SFHs) and proving a prolonged and multiepisode star formation activity.}
   {Leveraging ground- and space-based high-resolution images, we derived the SFH of Terzan~5 by employing the color-magnitude diagram fitting routine SFERA.}
   {The best-fit solution predicts an old, main peak occurred between 12 and $13\,$Gyr ago that generated $70\,\%$ of the current stellar mass, followed by a lower-rate star formation activity with two main additional bursts.}
   {These results indicate that Terzan 5, similarly to Liller 1,  experienced a prolonged, multiepisode star formation activity,  fueled by metal-enriched gas deposited in its central regions, in agreement with the expectations of a self-enrichment scenario in a primordial massive clump.}


   \maketitle
%

\section{Introduction}
\label{sec:intro}

Thanks to the photometric and spectroscopic campaigns that had been performed over the last twenty years, the picture of the 3D structure, chemistry and kinematics of the Galactic bulge has improved substantially, suggesting a composite formation process. However, no general consensus has been reached yet about the exact mechanisms that generated and shaped the bulge in the Milky Way and analogous galaxies.

Two main formation channels are currently invoked to reproduce the observed chemodynamical properties of the Milky Way's inner region: on the one hand, its barred and X-shaped morphology (\citealt{wegg+2013,wegg+2015}), together with a global cylindrical kinematics (\citealt{kunder+2012,ness+2013,zoccali+2014,rojasarriagada+2020}), would suggest a dynamical secular evolution of the Galactic disk driven by bar and buckling instabilities (\citealt{combes+1990,patsis+2002,athanassoula2005,shen+2010,fragkoudi+2017,debattista+2017}); on the other hand, the dominant old ages ($>10\,$Gyr, \citealt{clarkson+2008,clarkson+2011,valenti+2013,renzini+2018,surot+2019}) and the chemical pattern of its stars (e.g., \citealt{zoccali+2017,rojasarriagada+2017,rojasarriagada+2020,queiroz+2021}) requires an intense in-situ star formation occurred in a very short timescale  ($1-2\,$Gyr), as predicted by chemical evolution models (see \citealt{rojasarriagada+2017,matteucci2021}). 

One of the most promising ways to power this second formation path \-- a first and vigorous assembly phase of the bulge \-- is the merging of giant star-forming clumps generated by disk instabilities. Indeed, 
it has been theoretically shown that turbulent gas-rich disks can fragment into massive substructures  of the order of $10^7-10^9\,M_\odot$ (e.g., \citealt{behrendt+2016}). If long-lived enough (mainly determined by different feedback prescriptions), these massive structures can gravitationally interact, causing a redistribution of angular momentum outward and migration of clumps and gas inward. Numerical simulations demonstrate that this process forms a bulge in a characteristic timescale of $\sim1\,$Gyr for a Milky Way-like galaxy (\citealt{noguchi1999,immeli+2004a,immeli+2004b,bournaud+2007,carollo+2007,elmegreen+2008,dekel+2009}). Clumps developed by hydrodynamical simulations are also characterized by high star formation rate (SFR) densities (\citealt{clarke+2019}) and provide a natural way of reproducing the chemical trend displayed by bulge field stars (\citealt{debattista+2023}). 
Moreover, the clumpy morphology is a common feature of star-forming galaxies at cosmic-noon (see, e.g., \citealt{elmegreen+2009,guo+2015, shibuya+2016}).
The majority of the observed clumps substantially contribute to the rest-frame UV light of their host galaxies (up to $\sim40\%$), denoting very high SFRs (e.g., \citealt{zanella+2015,messa+2022}); however, a fraction of redder substructures is also detected (\citealt{guo+2015}). The measured stellar mass values range between $10^5$ and $10^9\,M_\odot$, with sizes from a few to hundred parsecs (\citealt{dessaugeszavadsky+2017,messa+2022,vanzella+2022a,vanzella+2022b,claeyssens+2023}). 

Yet, the actual contribution of high-redshift clumps to the bulge assembly is still a debated open question in both the galactic and extragalactic scientific communities. 
On the basis of current observations, it is very challenging to robustly evaluate if clumps survive long enough to migrate to the central region of the galaxy (e.g., \citealt{cava2018}). 
On the other hand, recent theoretical studies show that not only the massive clumps are able to migrate to the center, but some of them are also able to survive and to generate stellar clusters (\citealt{bournaud2016,dekel+2023}). 
It is likely that the primordial clumps were massive enough to retain iron-enriched supernova ejecta, creating a reservoir of gas to fuel, in a self-enrichment fashion, multiple bursts of star formation. This process would give rise to multi-iron and multi-age stellar subpopulations, at odds with what observed in genuine globular clusters (GCs). 
These in-situ stellar systems would be the empirical witnesses of the hierarchical assembly of the Galactic bulge, and are therefore called Bulge Fossil Fragments (BFFs, \citealt{ferraro+2009,ferraro+2021}).
Thus, the detection of BFFs in the bulge would represent a key observational evidence in support to the involvement of clumps in building bulges, which is fundamental for our understanding of the formation of Milky Way-like galaxies in a broad cosmological context (i.e., to guide the formation of these galaxies implemented in current cosmological simulations; to have constraints on the gas fraction, turbulence and kinematics of high-redshift disk galaxies; to understand the feedback process prescriptions; etc.).

To date, only two BFF candidates have been detected in the bulge: Terzan~5 (\citealt{ferraro+2009,ferraro+2016}) and Liller~1 (\citealt{ferraro+2021}). As suggested by isochrone fits, these massive (few $10^6\,M_\odot$, \citealt{lanzoni+2010,saracino+2015}) stellar systems show a color-magnitude diagram (CMD) simultaneously populated by old ($\sim12\,$Gyr) and young (up to $1\,$Gyr) stellar populations. In addition, spectroscopic analyses of these systems demonstrate a tight chemical similarity with bulge field stars. Indeed, their old stars are metal-rich ($\rm{[Fe/H]}\ge-0.5\,$dex) and $\alpha$-enhanced ($\rm{[\alpha/Fe]}\sim +0.3\,$dex), while young stars are characterized by super-solar metallicity ($\rm{[Fe/H]}\sim + 0.3\,$dex) and solar-scaled [$\alpha$/Fe] abundance ratio (\citealt{origlia+2011,origlia+2013,origlia+2019,massari+2014,crociati+2023,alvarezgaray+2024}). Moreover, recent chemical evolutionary models specifically computed for the case of Terzan~5 (\citealt{romano+2023}) nicely reproduce all the chemical patterns observed in this stellar system in the framework of the self-enrichment evolution of a progenitor with an initial mass of $4\times10^7\,M_{\odot}$. However, alternative formation models have been proposed for the origin of these two stellar systems. In particular, \citet{mckenzie+2018} and \citet{bastian+2022} suggested that the young stellar component was originated from the accretion of a field Giant Molecular Clouds (GMCs) by a genuine GC Indeed, according to current numerical simulations (\citealt{mckenzie+2018}), these should be extremely rare and fine-tuned events that would give rise to a single and brief star formation burst. Hence, this scenario would predict only two metallicity components and it would be clearly incompatible with a multiple burst and prolonged star formation history \citealt{bastian+2022}.

Determining the star formation histories (SFHs) of Terzan 5 and Liller 1 therefore represents a crucial step toward assessing their true origin. In the case of resolved stellar populations, a proper recovering of the SFH is obtained from synthetic CMD fitting methods (e.g., \citealt{tosi+1990,tolstoy+1996,dolphin1997,dolphin2002,hernandez+1999,cignoni&tosi2010,aparicio&hidalgo2009,weisz+2012,ruizlara+2018}). These aim to determine key properties  of composite stellar populations, such as age, mass, metallicity, and SFR, from the features readable in the CMD. Indeed, this analysis can unlock an amount of crucial information \-- namely, the duration of the star formation episodes, the presence of intermediate and barely detectable star formation bursts, and the build-up of the stellar mass throughout the lifetime of the systems \-- that cannot be obtained by simply performing isochrone fitting. 

\begin{figure}[!t]
    \centering
    \includegraphics[width=9cm]{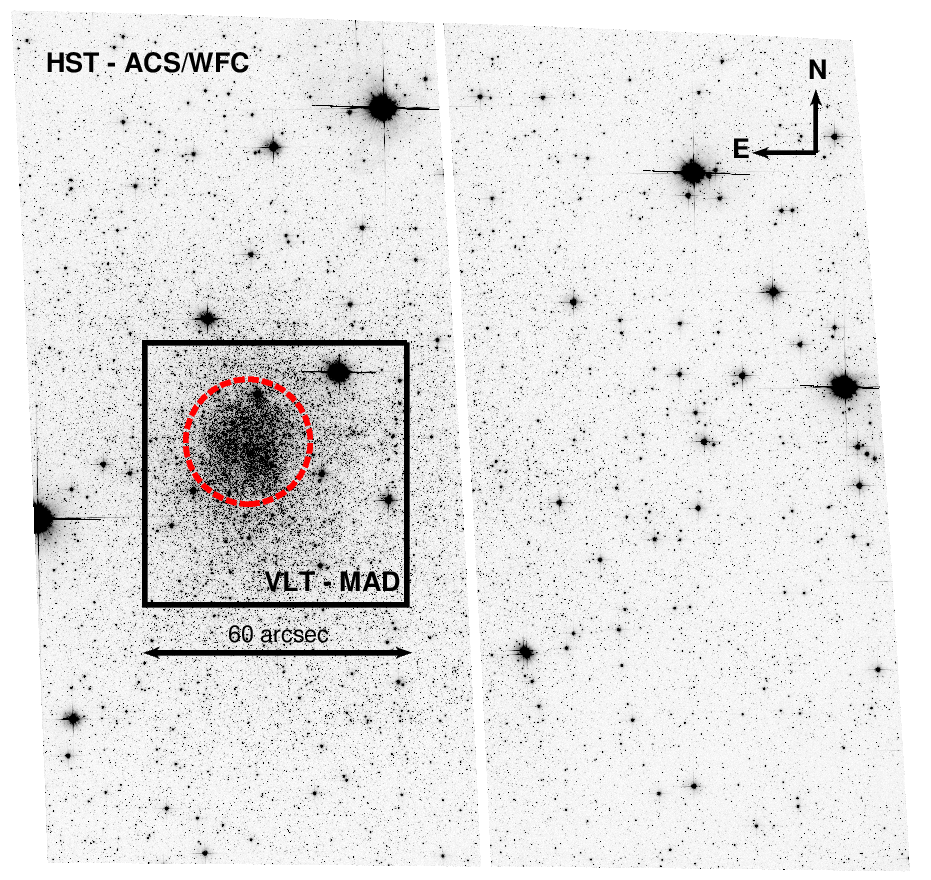}
    \caption{HST ACS/WFC image of Terzan~5 in the F814W filter. The field of view is $204\arcsec \times 204\arcsec$. The black box corresponds to the VLT-MAD pointing ($60\arcsec \times 60\arcsec$). The region used for the SFH reconstruction is beyond the red circle, corresponding to a radius of $15\arcsec$. North is up, east is to the left.}
\label{fig: ter5_sfh_fov}
\end{figure}

\begin{figure*}
    \centering
    \includegraphics[width=0.8\textwidth]{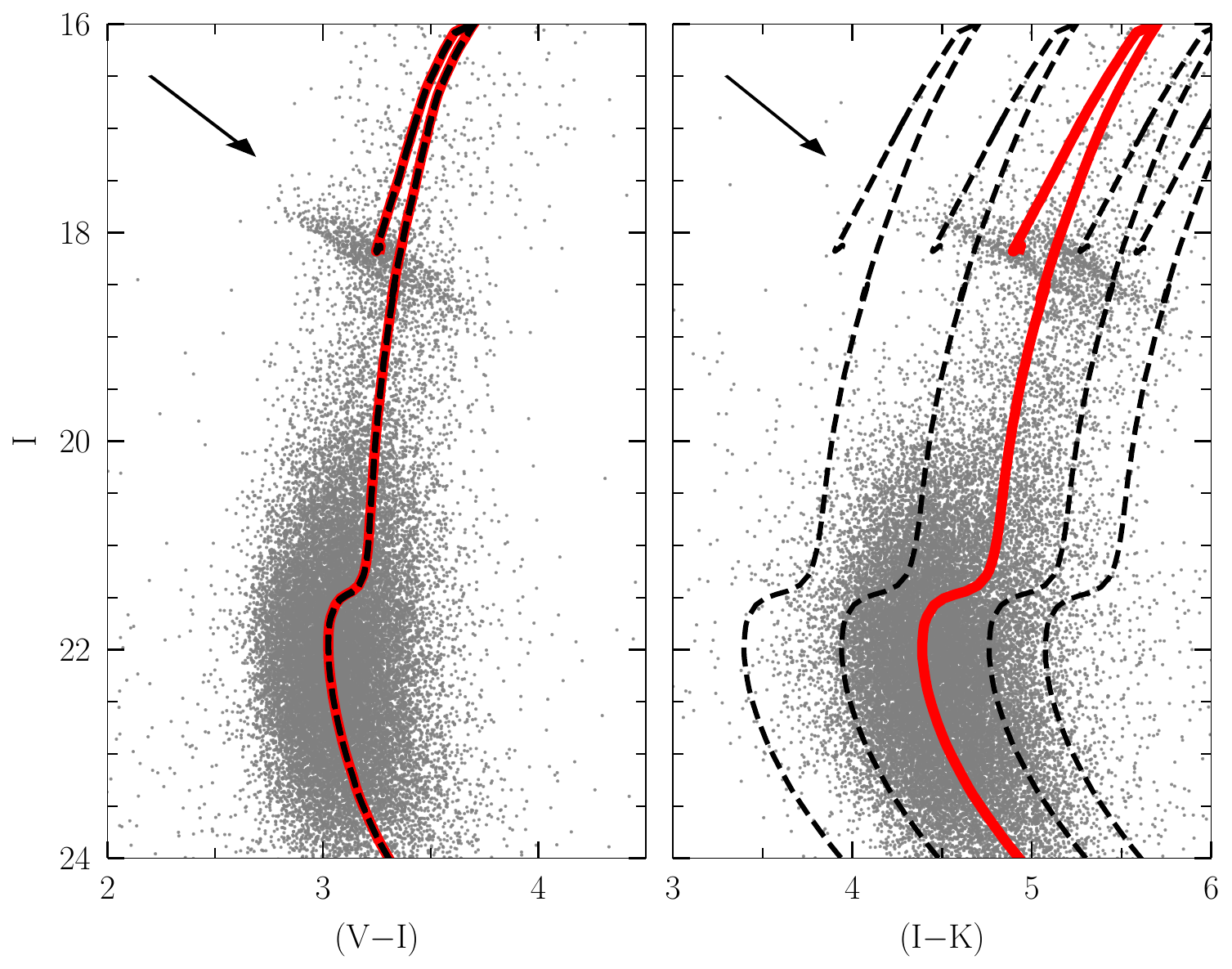}
    \caption{Optical (I, V$-$I) and hybrid (I, I$-$K) CMDs of Terzan~5 obtained through the combined HST and MAD observations, corresponding to the FOV delimited by the black box in Figure~\ref{fig: ter5_sfh_fov}. No quality cuts and differential reddening corrections are applied to these diagrams. It is worth noticing the distortion of the evolutionary sequences along the direction of the reddening vectors, indicated by the black arrows in the upper-left part of the panels. The red line corresponds to a $12\,$Gyr old isochrone (\citealt{bressan+2012, marigo+2017}) with $\text{[M/H]}= -0.2\,$dex, $E(B-V)=2.38$, and $\mu_{0}=13.87$ (reference parameters from the database of \citealt{valenti+2007}), reddenned by adopting $R_V=3.1$. The dashed black lines show the same isochrone shifted by assuming $R_V = 2.1, 2.6, 3.6, 4.1$ (from left to right). We varied $E(B-V)$ and $\mu_0$ to fit the optical (I, V$-$I) CMD. However, despite the remarkable broadening of the evolutionary sequences, these alternative $R_V$ values are clearly unable to match the hybrid CMD.}
\label{fig: ter5_sfh_cmd_noDRC}
\end{figure*}

\begin{figure}
    \centering
    \includegraphics[width=9cm]{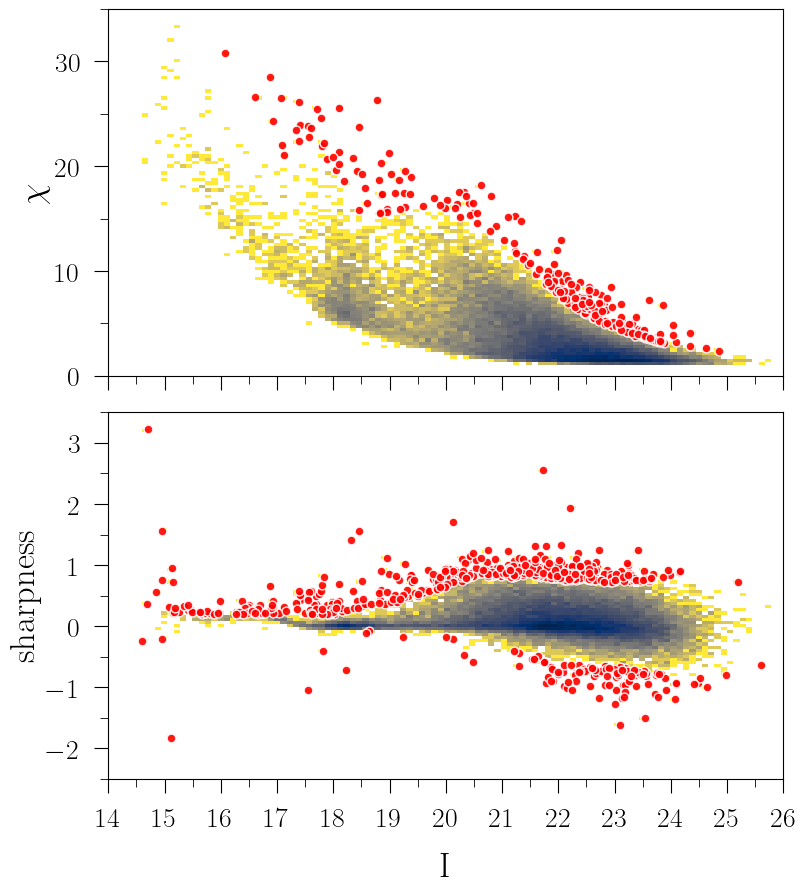}
    \caption{Trends of the photometric quality parameters $\chi$ and $sharpness$ as a function of the I magnitude resulting from the combined optical-IR reduction. Darker colors are associated to denser populated cells. Red points are stars excluded by the $3\sigma$ clipping procedure. }
\label{fig: ter5_sfh_photo}
\end{figure}

Within this context, in \cite{dalessandro+2022} we derived the SFH of Liller~1. Interestingly, we measured three main and prolonged ($\Delta t>1\,$Gyr) star formation events started at $\sim 13, 9$ and $3\,$Gyr ago, which generated the 75\%, 15\% and 10\% of the system's mass, respectively. These results are in line with those expected from the BFF scenario, reinforcing the hypothesis that Liller~1 is the surviving remnant of a much more massive primordial structure that experienced self-enrichment and likely contributed to the Galactic bulge formation.
Here we present an analogous analysis on Terzan~5. Previous age-dating studies based on isochrone fitting (\citealt{ferraro+2016}) allowed the detection of an old and a young main sequence turn off (MS-TO), measuring an age of $12\pm1\,$Gyr and $4.5\pm0.5\,$Gyr, respectively. In fact, as in the case of Liller 1, to properly assess the total number of bursts and their time duration, as well as the overall efficiency of the star formation activity experienced by the system it is necessary to reconstruct the entire SFH.
The manuscript is structured as follows: in Section~\ref{sec: input_catalog} we describe all the steps we performed to obtain the observed CMD of Terzan~5, including the photometric reduction, the correction for the effects of differential reddening and the evaluation of the observational uncertainties; in Section~\ref{sec: SFH} we present the adopted method for reconstructing the SFH and the main outcomes; in Section~\ref{sec: summary_conclusions} we summarize and critically evaluate our results.


\section{The input photometric catalog}
\label{sec: input_catalog}

\subsection{Observations and data reduction}
\label{sec: ter5_sfh_reduction}

As demonstrated in \cite{dalessandro+2022}, the combination of high spatial resolution infrared (IR) and optical photometric observations is a successful choice for analyzing the SFH of heavily extincted stellar clusters like Liller~1 (\citealt{pallanca+2021a}). Consequently, the data set employed in this study \-- focused on Liller~1's sibling, Terzan~5 \-- is a combination of archival images covering both optical and IR bands.

Specifically, the IR images were acquired with the Multi-conjugate Adaptive Optics Demonstrator (MAD, science demonstration proposal, PI: Ferraro), which had been installed for a brief trial period at the ESO/VLT.
This data set is composed of 15 exposures in the K filter\footnote{Throughout the manuscript, we refer to the filters $K_{s}$, F606W and F814W with the labels K, V and I, respectively.} roughly centered on Terzan~5, with a dither pattern resulting in a final field of view (FOV) of 1$\arcmin\times1\arcmin$ (see the black square in Figure~\ref{fig: ter5_sfh_fov}). The IR images used on this work are a subsample  
selected on the basis of the seeing quality, i.e. with a stable FWHM in the K-band of $0.1\arcsec$ across the entire MAD FOV. The integration time of each exposure is $t_{exp} = 120\,$s. 
The optical imagery consists of 10 exposures in the filters V and I acquired with the ACS/WFC on board the Hubble Space Telescope (HST), secured through proposal GO 12933 (PI: Ferraro). All the exposures are $t_{exp}= 365\,$s long.

The aforementioned images have been already utilized in previous photometric studies of Terzan~5 \citep[e.g.,][]{massari+2015, ferraro+2016}.
However, the reconstruction of the SFH from synthetic CMD fitting methods requires not only large samples of resolved stars, but also high photometric accuracy, the proper assessment of photometric completeness, and accurate corrections for differential reddening. Hence, following the same strategy adopted in the case if Liller 1 (see \citealt{ferraro+2021, pallanca+2021a,dalessandro+2022}), we reanalyzed those images exploiting simultaneously the information coming from all the available exposures. This approach has been shown to lead to an increase in the statistics, photometric accuracy and the depth. Moreover, as discussed in Section~\ref{sec: ter5_sfh_as}, it allowed us to precisely reconstruct all the steps of the photometric reduction, which is necessary to estimate the photometric incompleteness and uncertainties. In tunr, this also brought to an improved estimate of the differential reddening affecting the system (see Sect. \ref{sec: ter5_sfh_reddening}).

The photometric reduction was performed through PSF fitting by using the software \texttt{DAOPHOTIV} \citep{stetson1987}. First, PSF models were determined via the \texttt{PSF} routine for each exposure, 
selecting $\sim200$ isolated, bright and not saturated stars homogeneously distributed across the entire FOVs. The analytical PSF functions were complemented by a spatially variable lookup table, following a cubic polynomial spatial variation. We applied the best PSF models to all the sources corresponding to flux peaks at least $3\sigma$ above the local background by using the \texttt{ALLSTAR} routine. 

Differently from previous photometric analyses of Terzan~5, these first catalogs of instrumental magnitudes and positions were combined together via the cross-matching subroutines \texttt{DAOMATCH} and \texttt{DAOMASTER} (\citealt{stetson1993}), generating an input master starlist for \texttt{ALLFRAME} \citep{stetson1994} which includes the information coming from both optical and IR images. Specifically, the master list was generated by including the sources detected in more than five images in the K filter, and in more than three images in at least one of the optical ones. The strength of this analysis is that stars detected only in the optical images were then searched and analyzed also in the K images (and vice versa), because the \texttt{ALLFRAME} routine forces a fit at the corresponding positions of stars in
the master list. The information provided by the two filters separately is thus maximized, with fruitful results in terms of photometric accuracy and statistics.

The final raw magnitude for each recovered star (in a specific filter) is the weighted mean of multiple magnitude estimates (one for each exposure) once homogenized via \texttt{DAOMATCH} and \texttt{DAOMASTER}, while the photometric error is the corresponding standard deviation. The final catalog includes 30,834 objects that have both I and K magnitudes. Therefore, it is limited to the overlapping region between the ACS/WFC and the MAD FOVs (see Figure~\ref{fig: ter5_sfh_fov}).

We calibrated the optical instrumental magnitudes into the HST VEGAMAG system by applying the recipes and time-dependent ACS zero points provided by the STScI web interface\footnote{\url{https://www.stsci.edu/hst/instrumentation/acs}}. Aperture corrections were determined with respect to a radius of 8 pixel, selecting a subsample of bright and isolated stars across the FOV, and referring to the value of Encircle Energy tabulated in the HST website. The IR raw magnitudes were calibrated onto the 2MASS photometric system by computing the offset of the common stars between our catalog and the public available VISTA Variables in the V\'ia L\'actea (VVV) survey (\citealt{minniti+2010}; \citealt{surot+2019}). Among the stars in common, we retained from the VVV catalog the ones with low photometric error ($\sigma_{\rm K}<0.075$), and we adopted as zero-point the median of the difference after a $3\sigma$ rejection. The resulting CMDs, both in optical and hybrid (I, I$-$K) set of filters, are shown in Figure~\ref{fig: ter5_sfh_cmd_noDRC}. The distributions of \texttt{DAOPHOT} photometric parameters $\chi$ and $sharpness$, which are related to the quality of the PSF fitting, are displayed in Figure~\ref{fig: ter5_sfh_photo}, and they will be used to clean the CMDs from spurious objects or badly fitted stars.

Since the final catalog was built considering one optical exposure as reference frame, we corrected the instrumental coordinates for the ACS/WFC geometric distortions following the prescription of \cite{anderson+2010}, and subsequently we cross-matched our catalog with the one from \cite{ferraro+2016} to retrieve HST-based high-resolution PMs calculated by \cite{massari+2015} and absolute coordinates $(\alpha, \delta)$ based on the astrometric 2MASS catalog. 

\begin{figure*}
    \centering
    \includegraphics[width=1\textwidth]{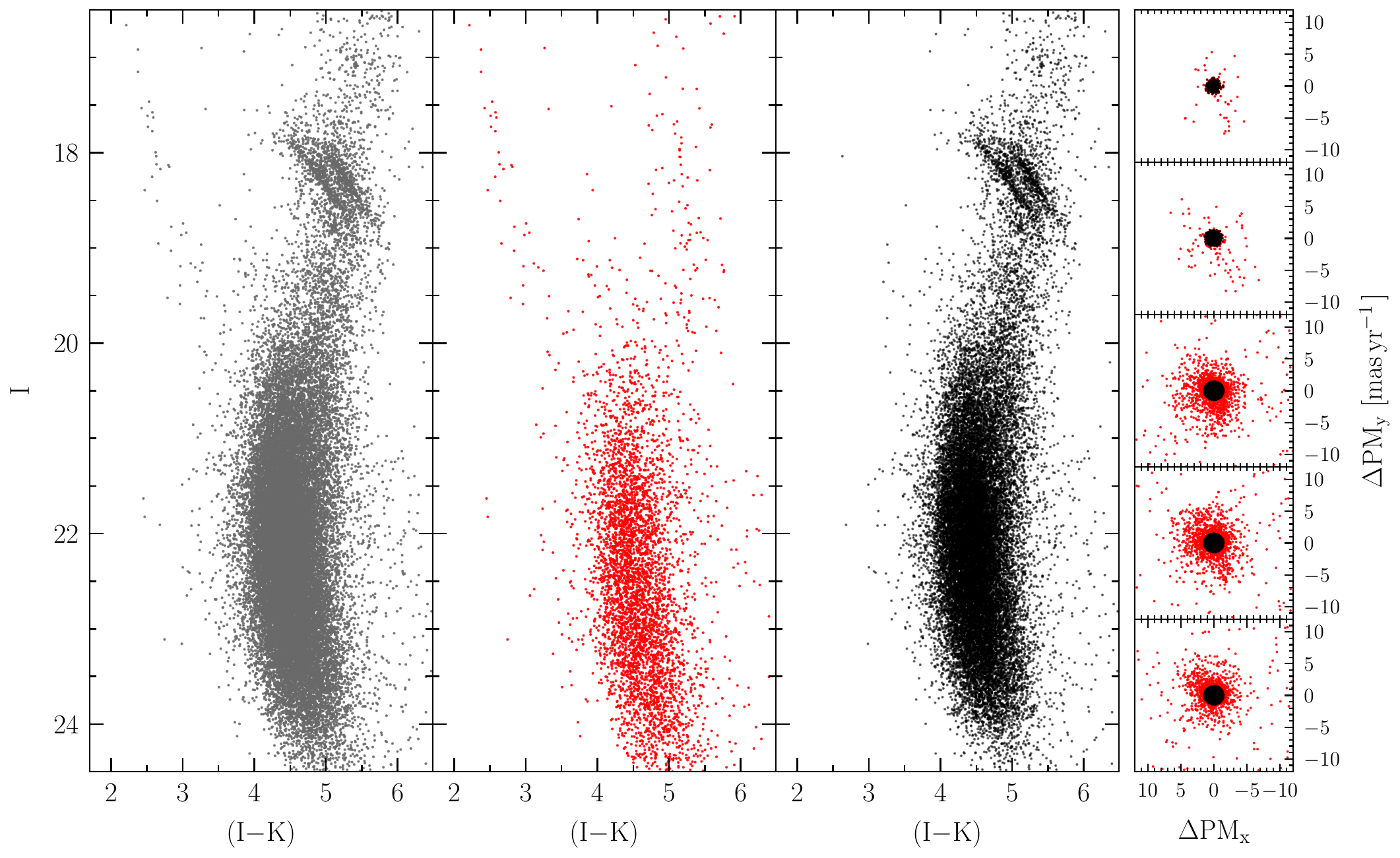}
    \caption{Cleaning procedure from field interlopers via PMs. \emph{Left panel}: CMD of the stars included in the MAD footprint with measured PMs from \cite{massari+2015}. \emph{Middle panel}: CMD of the field population as selected from the VPDs, that is stars with PMs not compatible with that of Terzan~5. \emph{Right panel}: PM-cleaned CMD of Terzan~5 obtained by considering only likely cluster members selected from the VPDs. \emph{Rightmost column}: VPDs of the measured stars divided in 5 bins of 1.5 mag each, starting at $\rm{I}_{\rm DRC}=16.5$. Black dots represent stars within a radius of $1.5\,\rm mas \,\rm yr^{-1}$ and correspond to likely cluster members according to the criterion of \cite{massari+2015}.} 
\label{fig: ter5_sfh_pm}
\end{figure*}

\subsection{Field contamination}
\label{sec: field_contamination}

The next step is to distinguish genuine member stars of Terzan~5 from Galactic field interlopers. To this purpose, we used the relative PMs determined by \cite{massari+2015}.The result of this procedure is shown in Figure~\ref{fig: ter5_sfh_pm}. The leftmost panel shows the CMD of all the stars observed within the MAD FOV. Their measured PMs are plotted in the vector point diagrams (VPDs) shown in the rightmost panel of the figure. Following \cite{massari+2015}, we flagged as members of Terzan 5 all the stars included within a radius of $1.5\,\text{mas}\,\text{yr}^{-1}$ from the origin of the VPD (black dots), while field interlopers are those beyond this radius (red dots). Their respective CMDs are shown in the third and second panels of Figure~\ref{fig: ter5_sfh_pm}, respectively.

Since the PM distribution of bulge field stars substantially overlap with the population of Terzan 5 in the VPD (see the rightmost panel of Fig. \ref{fig: ter5_sfh_pm}), the adopted criterion for the selection of Terzan 5 members (as those with PM $<1.5\,\text{mas}\,\text{yr}^{-1}$) necessarily implies some residual field contamination. To ensure that the recovered SFH will not be affected by bulge or disk interlopers, we thus estimated the fraction of residual field interlopers by using two different methods. Firstly, we used a Monte Carlo approach: we fitted the two PM components ($\Delta \rm PM_x$, $\Delta \rm PM_y$) observed in the VPD (black and red points in the rightmost panel of Figure~\ref{fig: ter5_sfh_pm}) with Gaussian functions, and subsequently we run $1000$ realizations of such Gaussians, each time counting the number field stars falling within the $1.5\,\text{mas}\,\text{yr}^{-1}$ radial limit. The contamination fraction is computed as the mean over the $1000$ realizations. Secondly, we utilized the Python package \texttt{scikit-learn} (\citealt{pedregosa+2011}) to exploit the Gaussian mixture model (GMM) statistics. We forced the algorithm to fit two Gaussian components to the distribution of points observed in the VPDs and we looked at the weight of the field stars component. In both cases, we obtained a contamination fraction lower than $5\%$, which is not expected to heavily influence the SFH analysis. The low fraction of field objects is somehow expected, since we are considering a small field of view centered on Terzan~5.

\subsection{High spatial resolution differential reddening map}
\label{sec: ter5_sfh_reddening}

The Galactic bulge is one of the most extincted region of the Milky Way. The large amount of interstellar dust causes the scattering and the absorption of the (blue) light coming from the stars, which appear redder and fainter with respect to their intrinsic colors and magnitudes, depending on the degree of heterogeneity of the medium across the observed FOV. This effect is clearly visible in Figure~\ref{fig: ter5_sfh_cmd_noDRC}, where all the evolutionary sequences of the CMD result blurred and elongated along the direction of the reddening vector (see the black arrows in Figure~\ref{fig: ter5_sfh_cmd_noDRC}). Given the high extinction in the direction of Terzan~5 and its location within the Galaxy ($E(B-V) = 2.38$; $l,b= 3.81^\circ, 1.67^\circ$; \citealt{valenti+2007}), this problem is exacerbated, and the correction for differential reddening effects becomes a crucial step in the identification and characterization of its subpopulations. Indeed, the study of \cite{ferraro+2016} was based on the differential extinction map presented in \cite{massari+2012}, which reached a spatial resolution of $8\arcsec\times 8\arcsec$. This resolution provided an overall good correction for differential reddening effects, but it led the detection of a double MS-TO only after the selection of the least extincted region of the MAD FOV, that corresponds to an area of $\sim 25\arcsec \times 25\arcsec$ in the south-west direction. To improve the accuracy of the correction, we decided to take advantage of the newly performed photometric analysis for building a higher resolution differential reddening map in the direction of Terzan~5.

It has been extensively shown that, when constructing differential reddening maps of stellar clusters, techniques based on the so-called star-by-star method can provide maps with a spatial scale of few arcseconds. Thanks to the improved statistics and photometric accuracy, we could model the differential reddening in the field of view covered by our observations with this novel approach, following the guidelines provided in several recent papers (e.g., \citealt{dalessandro+2018,saracino+2019,pallanca+2021a,cadelano+2020} and references therein). The star-by-star method estimates the differential reddening of each star in the catalog by computing the shift, along the reddening vector, needed to match the mean ridge line (MRL) of the cluster to the "local" CMD. The latter is defined as the CMD built with the objects spatially close to the targeted star (see below).

The direction and the components of the reddening vector\footnote{Specifically for the hybrid CMD, the components of the reddening vector are $V_x=A_I- A_K$; $V_y=A_I$.} are fixed once a precise extinction law (i.e., the  dependence of the absorption coefficient $A$ on wavelength  $\lambda$) is adopted. Indeed, extinction laws are parameterized as (see e.g. \citealt{cardelli+1989})
\begin{equation}
    A_{\lambda} = R_V \times c_{\lambda, R_V} \times E(B-V), 
\label{eq: extinction_law}
\end{equation}
where $E(B-V)$ is the color excess, that is defined as the difference between the observed $(B-V)$ and the intrinsic $(B-V)_0$ color. The coefficient $c_{\lambda, R_V}$ represents the ratio $A_\lambda/A_V$, and it depends on the assumed extinction law (i.e., \citealt{cardelli+1989}, \citealt{odonnell1994} or \citealt{fitzpatrick1999}), the adopted value of the $R_V$ parameter, and the effective wavelength of the filter. It is conventionally set equal to $1$ in the (Johnson) $V$ band. $R_V$ is usually set $R_V = 3.1$, following fundamental studies about diffuse Galactic interstellar medium (\citealp{sneden+1978}). However, it has long been known that the coefficient $R_V$ can vary depending on the observed region of the Milky Way, particularly in the direction of the bulge (see, e.g., 
\citealt{johnson&borgman1963}; \citealt{johnson1965}; \citealt{frogel+1995}; \citealt{popowski2000}; \citealt{alonsoGarcia+2011}; \citealt{nataf+2013}; \citealt{casagrande+2014}; \citealt{alonsoGarcia+2017}; \citealt{saha+2019}; \citealt{pallanca+2021a}; \citealt{legnardi+2023}). A powerful tool to determine $R_V$ in the direction of a stellar system, even in the case of strong stretching of the evolutionary sequences, is to search for the value that allows the simultaneous match of the optical, NIR and hybrid CMDs with an isochrone of appropriate metallicity and age \citep[see, e.g., the case of Liller 1 in][]{pallanca+2021a}. In fact, changing the value of $R_V$ modifies the value of $c_{\lambda, R_V}$, making the extinction law more or less steeply dependent on wavelength, especially in the optical regime. Hence, following \citet{pallanca+2021a}, we computed an isochrone (\citealt{bressan+2012,marigo+2017}) with age $t = 12\,$Gyr (\citealt{ferraro+2016}) and global metallicity $\text{[M/H]}=-0.2\,$dex (\citealt{valenti+2007}, \citealt{massari+2014}), and we searched for possible combinations of $R_V$, color excess $E(B-V)$, and distance modulus $\mu_0$ able to make the isochrone simultaneously matching the observed optical and in the hybrid CMDs. We varied $R_V$ by steps of 0.5 around the canonical value ($R_V=3.1$), since \cite{legnardi+2023} showed that, for heavily extincted stellar clusters, only large differences (e.g., $\Delta R_V\sim 0.5$) can affect the local reddening correction by using the star-by-star method. As shown in Figure~\ref{fig: ter5_sfh_cmd_noDRC}, the best solution is obtained by adopting the standard value $R_V=3.1$ (red solid line), together with the color excess and distance estimated by \citet{valenti+2007} for Terzan 5, namely $E(B-V)=2.38$ and $\mu_0=13.87$. Therefore, we adopted the extinction coefficients calculated considering $R_V=3.1$, the extinction law from \cite{odonnell1994}, and effective wavelengths of the I (F814W) and K filters tabulated in \cite{rodrigo+2020}\footnote{$\lambda_{eff, \text{F814W}}= 7973.39\,\AA$ ; $\lambda_{eff, \text{K}}= 21590\, \AA$. We emphasize that also \citet{massari+2012} adopted $R_V=3.1$ in the Johnson V-band, while the value of 2.83 quoted in that paper refers to the F606W filter of the HST  ACS/WFC.}. Under these assumptions, the extinction coefficients are $R_{\text{I}}=1.90$ and $R_{\text{K}}=0.36$\footnote{We calculated the extinction coefficients by means of the Python package \texttt{extinction} (\citealt{barbary2016}) adopting the \cite{odonnell1994} extinction law.}

\begin{figure}
    \centering
    \includegraphics[width=9cm]{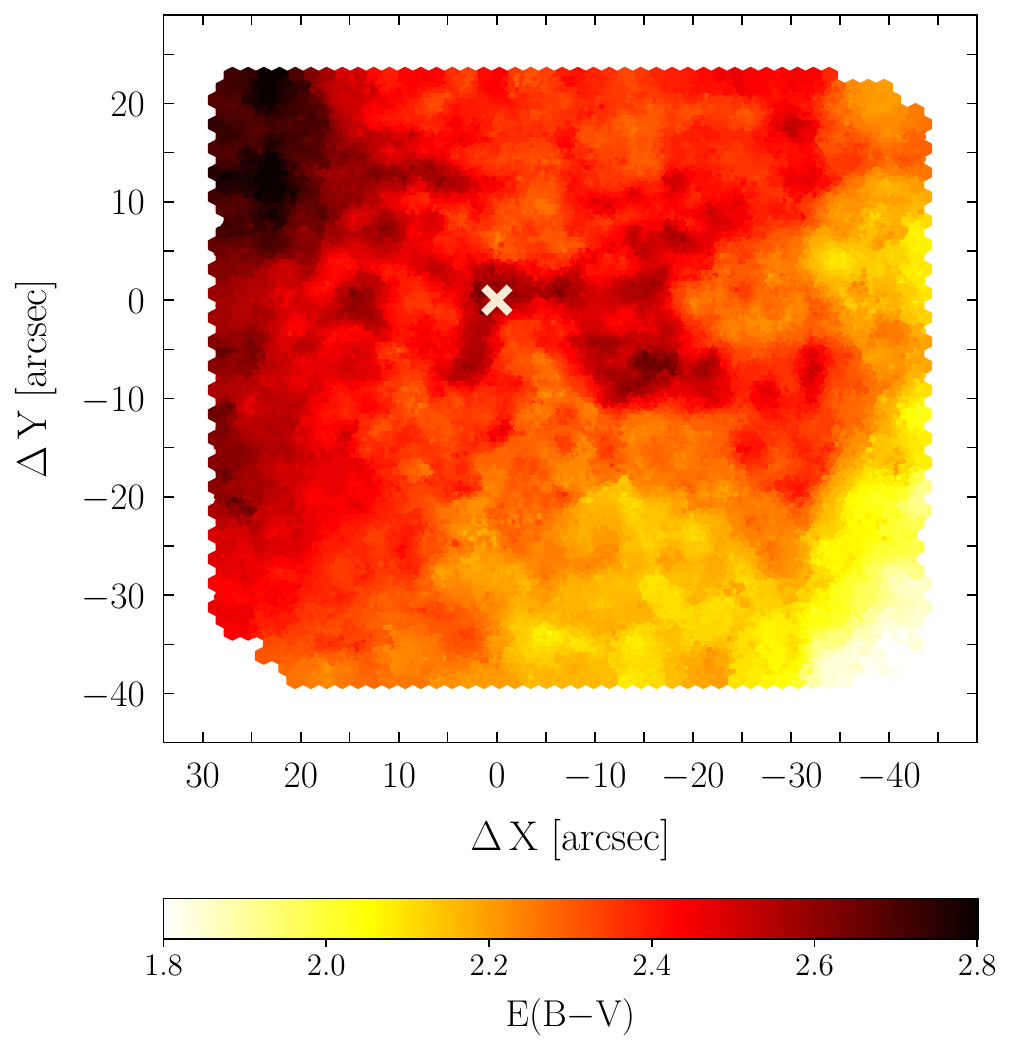}
    \caption{Map of the absolute reddening $E(B-V)$ in the direction of Terzan~5. Darker colors indicate the more extincted areas within the observed FOV. The white cross marks the gravity center of Terzan~5 (\citealt{lanzoni+2010}). North is up, while east is on the left. }
\label{fig: ter5_sfh_redd_mag}
\end{figure}

\begin{figure*}
    \centering
    \includegraphics[scale=0.45]{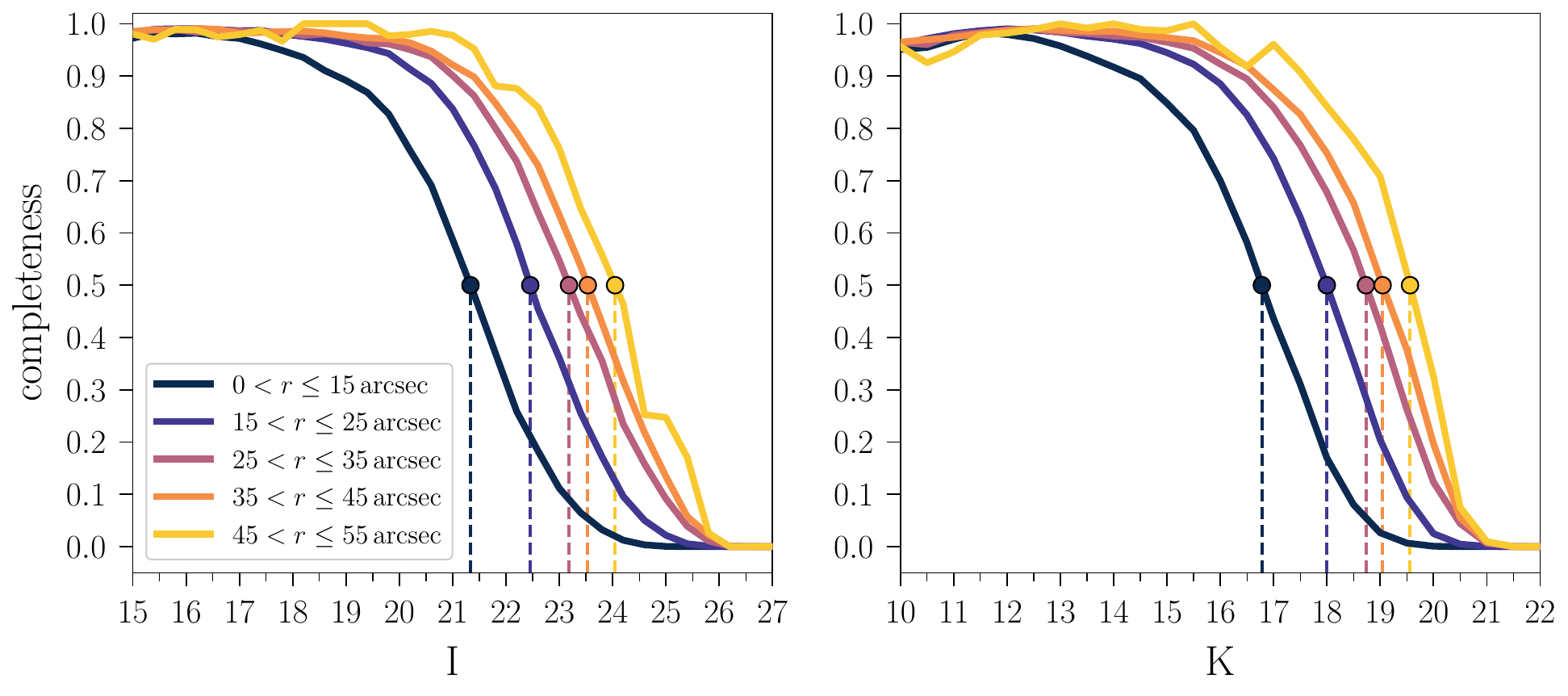}
    \includegraphics[scale=0.38]{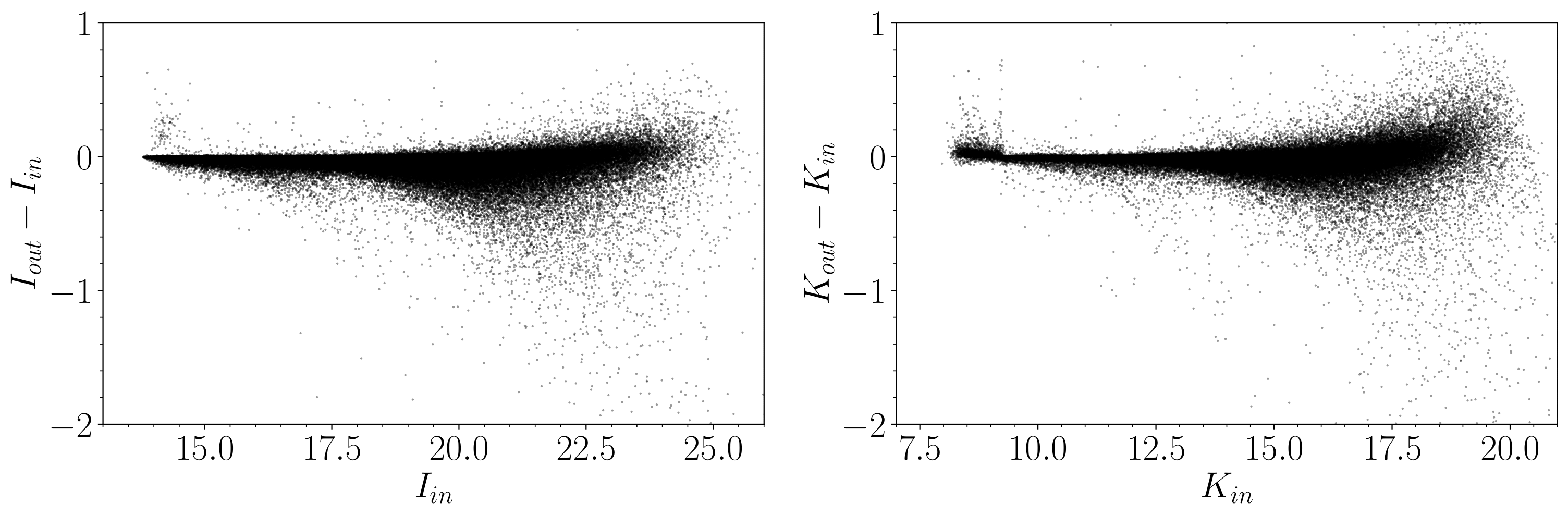}

    \caption{\emph{Upper panels:} completeness curves as a function of the I and K magnitudes in different concentric annuli from the cluster's center. The colored dots and vertical lines highlight the limit magnitude at which the $50\%$ is reached within the respective radial annulus. \emph{Lower panels:} photometric errors, i.e. the difference between the input and output magnitudes, in I and K bands recovered from the artificial star tests. }
\label{fig: ter5_sfh_compl}
\end{figure*}

We then proceeded to estimate the reddening of each star of the catalog, following the procedure extensively explained in \cite{pallanca+2021a}. The corrections were computed in the PM-cleaned (see Section~\ref{sec: field_contamination} for details) and hybrid (I, I$-$K) CMD.
Briefly, we first determined the MRL that approximates the upper main sequence, subgiant branch and red giant branch of Terzan~5. To do so, we divided the CMD in different magnitude bins adapting the bin size to the evolutionary sequence (from $0.2$ to $1$ mag). We calculated the mean values of color and magnitude of each bin by considering only a subsample of member high-quality stars, chosen through a $3\sigma$ rejection applied to the distributions of $\chi$ and $sharpness$ (see colored cells in Figure~\ref{fig: ter5_sfh_photo}).
The next step is to determine the value of differential reddening $\delta E(B-V)$ for all the stars. Specifically, for each star in the catalog, we first built the local CMD by using the $N_{\star}=30$ high-quality closest targets. Subsequently, the MRL is shifted along the direction of the reddening vector in steps of $\delta E(B-V)$, until it fits the local CMD. The value of reddening that minimizes the distance between the MRL and the local CMD is found by computing the residual color $\Delta\text{IK}$, defined as:

\begin{equation}
    \Delta\text{IK} = \sum_{i=1}^{N_{\star}} \vert \text{IK}_{\text{obs}, i} -  \text{IK}_{\text{MRL}, i} \vert + w_i \cdot  \vert \text{IK}_{\text{obs}, i} -  \text{IK}_{\text{MRL}, i} \vert \,\,\,,
\label{eq: redd1}
\end{equation}

where $w_i$ is a weight assigned to the stars of the local CMD, in order to valorize stars with small photometric errors or distances from the targeted star. It is defined as

\begin{equation}
    w_i = \frac{1}{d_i\cdot \sigma_i} \Biggl[ \sum_{j=1}^{N_{\star}} \Biggl( \frac{1}{d_j\cdot\sigma_j}\biggr) \biggr]^{-1} \,\,\,.
\label{eq:redd2}
\end{equation}

In Equations~(\ref{eq: redd1}) and (\ref{eq:redd2}), $\text{IK}_{\text{obs}, i}$ is the observed color of each star composing the local CMD,  $\text{IK}_{\text{MRL}, i}$ is the color of the MRL at the same magnitude level, $d$ is the projected distance from the selected star, and $\sigma$ the photometric uncertainty on the color. Finally, the assumed value of $\delta E(B-V)$ is the one that minimizes the normalized residual color ($\Delta\text{IK}/N_{\star}$). 

The resulting absolute differential reddening map in the direction of Terzan~5 (assuming the average color excess estimated in \citealt{valenti+2007}) is shown in Figure~\ref{fig: ter5_sfh_redd_mag}. This map has a typical spatial resolution of $1\arcsec-2\arcsec$ across the observed FOV, except for small regions limited to its very edge where the resolution is equal to $\sim3\arcsec$. It is worth noticing that, despite the small area covered by the IR camera ($60\arcsec \times 60\arcsec$), the maximum variation of $E(B-V)$ across the FOV is $\sim1$ mag, demonstrating that it is necessary to properly model the patchy absorption pattern of the dust clouds on small (arcsecond) scales. Indeed, the differential extinction map from \cite{massari+2012}, built with a larger cell size ($8\arcsec \times 8\arcsec$), quoted an extinction variation of $\delta E(B-V)=0.67\,$mag. The benefit from the differential reddening correction can be appreciated from the comparison between the observed CMD (Fig. 2) and the differential reddening corrected (DRC) CMD shown in the left-hand panel of Figure \ref{fig: cmd_drc}: indeed, independently of a slightly smaller color span along the x-axis, all the evolutionary sequences appear substantially less distorted along the reddening vector direction in Fig. \ref{fig: cmd_drc}.

\subsection{Artificial stars experiments}
\label{sec: ter5_sfh_as}

A new photometric analysis has been necessary also for the determination of the incompleteness and photometric errors of the input photometric catalog through artificial star tests. In fact, this method requires to re-perform the same photometric analysis that produced the final catalog once artificial stars are placed onto the images. 

Following the same procedure adopted in the case of Liller~1 (\citealt{dalessandro+2022}), we created a catalog of artificial stars extracting input magnitudes in the I band ($I_{in}$) from a flat luminosity function (LF). The extremes of the LF were chosen to match the observed CMD. We added the input artificial stars on the images with the PSF models and Poisson noise resulting from the photometric analysis thanks to the \texttt{DAOPHOT ADDSTAR} package. The stars were placed on the ACS/WFC images following the stellar density profile modeled by \cite{lanzoni+2010}, which corresponds to a King model (\citealt{king1966}) with structural parameters $c=1.49$, $r_{core}=9\arcsec$, and $r_t=4.6\arcmin$. To avoid overcrowding, we divided the images into cells of $20\times20\,\text{pixel}^2$ (more than 10 times the FWHM measured in the ACS/WFC), where only one artificial star was injected. We assigned to each artificial star a $K_{in}$ and a $V_{in}$ magnitude from a randomly extracted color in the range of the observed (I$-$K) and (V$-$I) colors. In this way, we could homogeneously cover the color-magnitude plane of the observed stars.
Thanks to the magnitudes homogenization and the geometrical transformation between the different detectors and exposures calculated by \texttt{DAOMATCH} and \texttt{DAOMASTER}, we ensured to place artificial stars in the same location in the V, I and K images, assigning the proper frame-dependent magnitude.
We then performed the same photometric analysis described in Section~\ref{sec: ter5_sfh_reduction} on the images containing artificial stars, and, at the end of the reduction, we cross-matched the output catalog with the one containing only input artificial stars. We iterated this procedure until the final catalog of artificial stars counted $\sim 700,000$ objects within the common FOV between the optical and IR data sets (see Figure~\ref{fig: ter5_sfh_fov}).

In this way, we could retrieve two important measurements: on the one hand, it is possible to count the fraction of stars lost during the photometric procedure, obtaining an estimate of the completeness of the photometric catalog as a function of color and magnitude; on the other hand, for the recovered stars, the difference between the input and the output magnitude provides a more realistic estimate of the photometric errors. Specifically, the completeness was defined as the ratio between the number of recovered stars and that of the injected ones. An  artificial star was considered "lost" when it is not recovered in the I or K filter, or when it is recovered more than 0.75 magnitude brighter. Indeed, in this latter case, the artificial star is likely blended with one or more other real stars, creating an "altered" source with more than twice the flux of the input one. This completeness is a sensitive function of the crowding conditions, as shown by the curves displayed in the upper panels of Figure~\ref{fig: ter5_sfh_compl}, built by dividing the FOV in annuli centered on the cluster center (from \citealt{lanzoni+2010}). As shown in Figure~\ref{fig: ter5_sfh_compl}, the completeness significantly drops in the first $15\arcsec$ from the cluster center. Within this distance bin, the magnitude corresponding to the 50\% completeness is $\text{I}=21.3$ (or $\text{K}=16.8$), which is $\sim0.3$ mag brighter than the MS-TO of the old population (assuming literature parameters, see the red isochrone in Figure~\ref{fig: ter5_sfh_cmd_noDRC}). On the other hand, the photometric completeness is equal to or better than 80\% at the old MS-TO level at larger radii. We concluded that the severe crowding of the innermost $15\arcsec$ of Terzan~5 irremediably hampers the use of the MS-TO for the determination of its SFH, and therefore we restricted the analysis to larger radial distances. The radial limit of $15\arcsec$ is marked by a dashed red circle in Figure~\ref{fig: ter5_sfh_fov}. Considering the region beyond this radial cut in the field of view covered by the observations, we still sample the $\sim45\%$ of the total mass of Terzan~5, computed from the best-fit King model of \cite{lanzoni+2010}.

Photometric errors and blends are assigned to the synthetic stars by using the distributions of $(I_{out}-I_{in})$ and $(K_{out}-K_{in})$ displayed in the bottom panels of Figure~\ref{fig: ter5_sfh_compl}. As expected, the distributions become broader and more asymmetric toward fainter magnitudes, indicating larger photometric uncertainties and blend effects. 

\section{Reconstruction of the star formation history}
\label{sec: SFH}

\subsection{Brief description of the method}
As for the case of Liller~1 (\citealt{dalessandro+2022}), we reconstructed the SFH exploiting the synthetic CMD fitting technique (e.g., \citealt{tosi+1990,tolstoy+1996,dolphin1997,dolphin2002,hernandez+1999,cignoni&tosi2010,aparicio&hidalgo2009,weisz+2012,ruizlara+2018}) through the code SFERA (Star Formation Evolution Recovery Algorithm, \citealt{cignoni+2015}).

SFERA parameterizes the SFH as a linear combination of $j \times k$  star formation events (also called basis functions, BFs), each characterized by a finite duration $\Delta t$ (centered at the time step $t_j$) and a narrow interval of metallicities $\Delta z$ (centered at the metallicity step $z_k$). For each synthetic star, we randomly extract age and metallicity from the intervals ($t_j$,$t_j + \Delta t$) and ($z_k$, $z_k + \Delta z$ ), and a stellar mass from the adopted initial mass function (IMF). Magnitudes are assigned by interpolating a library of isochrones. A percentage of synthetic stars is then coupled with a companion star, whose mass is extracted from the same IMF and whose flux is added to the primary one. To overcome the short duration of some evolutionary phases, the BFs are typically populated with a high number of synthetic stars (of the order of $10^6$). 

To produce realistic simulations, theoretical photometry is degraded to mimic the observational conditions of the data. This translates into including extinction and distance effects, and convolving the synthetic photometry with incompleteness and photometric errors. These observational uncertainties are quantified by means of extensive artificial stars experiments (see Section~\ref{sec: ter5_sfh_as} for details). Finally, BFs’ CMDs and the observational CMD are binned in n bins of color and m bins of magnitude to construct density maps (Hess diagrams). The result is a library of $j \times k$  Hess diagrams  $BF_{m,n} (j, k)$, whose elements are linearly combined to match the observational counterpart, following the equation below:

\begin{equation}
\label{eq: bf_linear_comb}
    N_{n,m} = \sum_{j}\sum_{k} a(j,k)\times \text{BF}_{m,n}(j,k)  \,\,\, .
\end{equation}
The coefficients $a(j, k)$, representing the SFR at the time step $j$ and metallicity step $k$, are found by minimizing a Poissonian likelihood:

\begin{equation}
\label{eq: poisson_likely_sfera}
    \chi^{2}_{P} = 2 \sum_{i=1}^{N_{bin}} e_{i} - o_{i} + o_{i}\ln \frac{o_{i}}{e{i}}  \,\,\,,
\end{equation}

where $e_i$ and $o_i$ are, respectively, the expected and the measured number of stars contained in the $i\text{th}$ cell of the binned CMDs. The uncertainties around the best-fit solution are estimated by bootstrapping the data and re-deriving the solutions.

From the minimization process we obtain the set of weights $a(j, k)$ from which, together with the total mass $M(j,k)$  and duration of each BF, we find the most likely SFH for the data set under analysis. Moreover, the mean metallicity of the "best" model in age step $t_j$ can be obtained by adding up all the possible metallicity values covered by the BF $z_k$, weighted by the ratio between their masses and the total mass of the "best" model in the corresponding age range: 

\begin{equation}
    z(t_j) = \sum_k z_k \frac{a(j,k) M(j,k)}{\sum_k a(j,k) M(j,k)}  \,\,\,.
\end{equation}
The procedure can be repeated over all the chosen age steps, in order to obtain an estimate of the "true" age-metallicity relation.

To explore the wide parameter space resulting from the choice of the age and metallicity steps, SFERA combines the genetic algorithm (GA) Pikaia\footnote{This is a public available routine developed at the High Altitude Observatory.} with a local search routine. On the one hand, the GA allows to explore the parameter space in more points simultaneously, being less sensitive to the initial conditions and local minima; on the other hand, the local search speeds up the convergence time and increases the accuracy of the solution (see \citealt{cignoni+2015} for an extensive and detailed description).

\begin{figure}
    \centering
    \includegraphics[width=9cm]{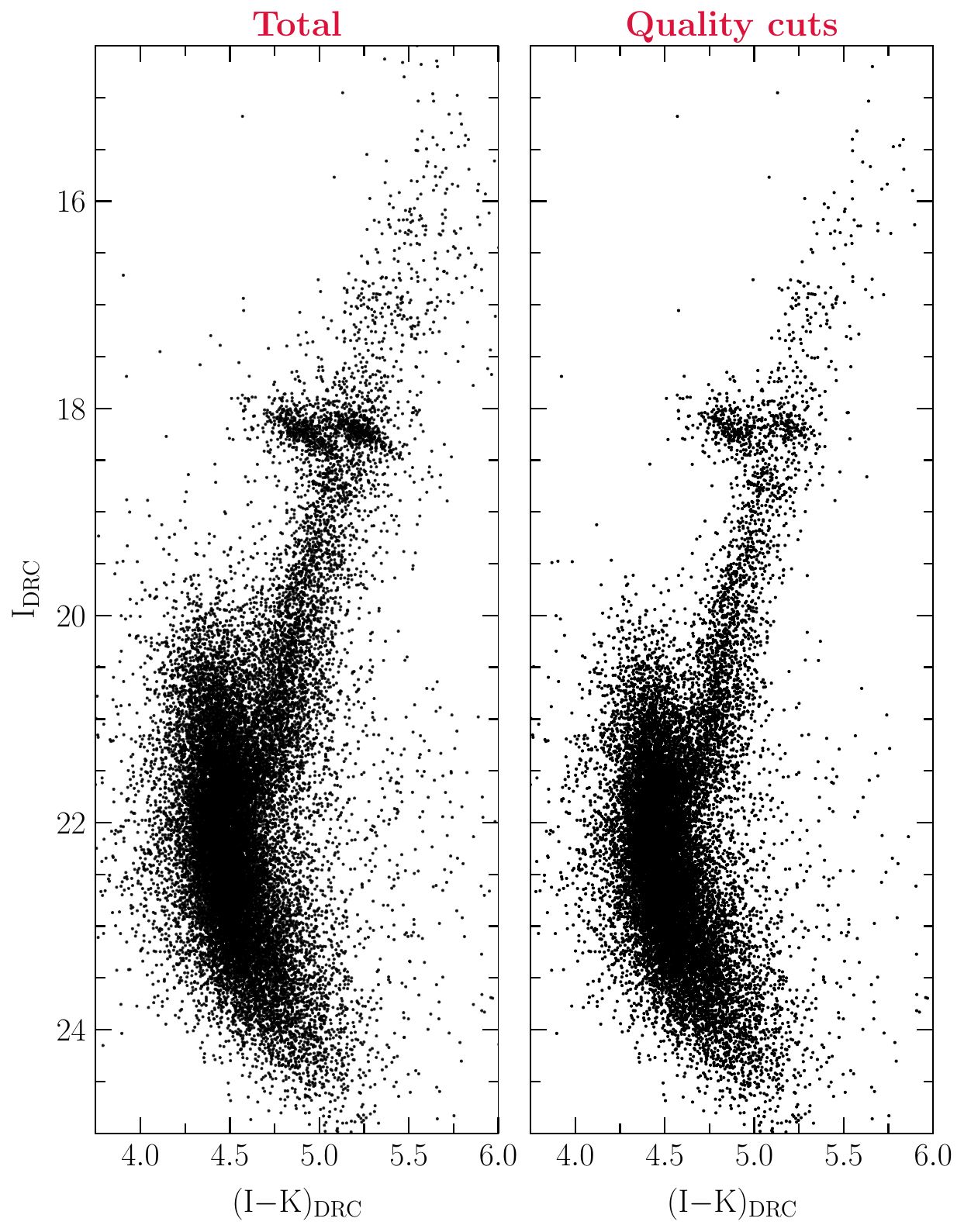}
    \caption{DRC and PM-cleaned CMDs of Terzan~5 obtained through the photometric reduction and the differential reddening map described in this work. The left-hand panel displays all the member stars measured in the FOV, while the right-hand panel shows the subsample of stars satisfying our quality cuts, specifically a radial distance of $r\ge15\arcsec$ from the center and good values of $\chi$ and $sharpness$ (i.e., within $3\sigma$ of their distributions; see Fig.~\ref{fig: ter5_sfh_photo}).}
    \label{fig: cmd_drc}
\end{figure}

\subsection{Input parameters}
\label{sec: sfera_input}
The BF CMDs used for the comparison with the observed one were built from the PARSEC-COLIBRI stellar isochrones (\citealt{bressan+2012,marigo+2017}), covering an age range from $2\,\text{Myr}$ to $13.4\,\text{Gyr}$, with a logarithmic step of $\Delta \log(t/\text{yr})=0.005\,$dex, and a metallicity range from $\text{[M/H]}=-1.5\,$ to $\text{[M/H]}=+0.3\,$dex\footnote{We assume $\rm{[M/H]}=\log(Z/Z_{\odot})$, with $Z_{\odot}=0.0152$.}, with a step of $0.025\,$dex. The isochrones were computed in the HST/ACS-WFC and 2MASS photometric systems, since these are the photometric bands used to build the observed CMD. 

The SFH is parameterized by 14 contiguous time steps of variable duration: 0.5 Gyr for stars younger than 1 Gyr, 1 Gyr for stars between 1 Gyr and 13 Gyr, 1.4 Gyr for older star. The metallicity step is fixed at 0.1 dex.
The BF CMDs  were generated  with a \citet{kroupa2001} IMF between 0.1 and 300~$M_{\odot}$. $30\%$ of synthetic stars are coupled with a stellar companion sampled from the same IMF to mimic unresolved binaries. 

The distance modulus ($\mu_0$) and average reddening $E(B-V)$ are initially adopted from the literature (\citealt{valenti+2007}), then the code is permitted to adjust these parameters to maximize the likelihood. In order to model the residual differential reddening, we added to each synthetic star a Gaussian reddening whose width is selected during the fitting process.

The CMD bin size is chosen as a compromise between the photometric error, resolution of the selected stellar models, and computational time, ranging from $0.1$ to $0.05$ mag both in magnitude and color. Finally, the routine does not assume any age-metallicity relation a priori.

\subsection{SFERA best-fit solution}
\label{sec: ter5_sfh_best}

\begin{figure}
    \centering
    \includegraphics[width=9cm]{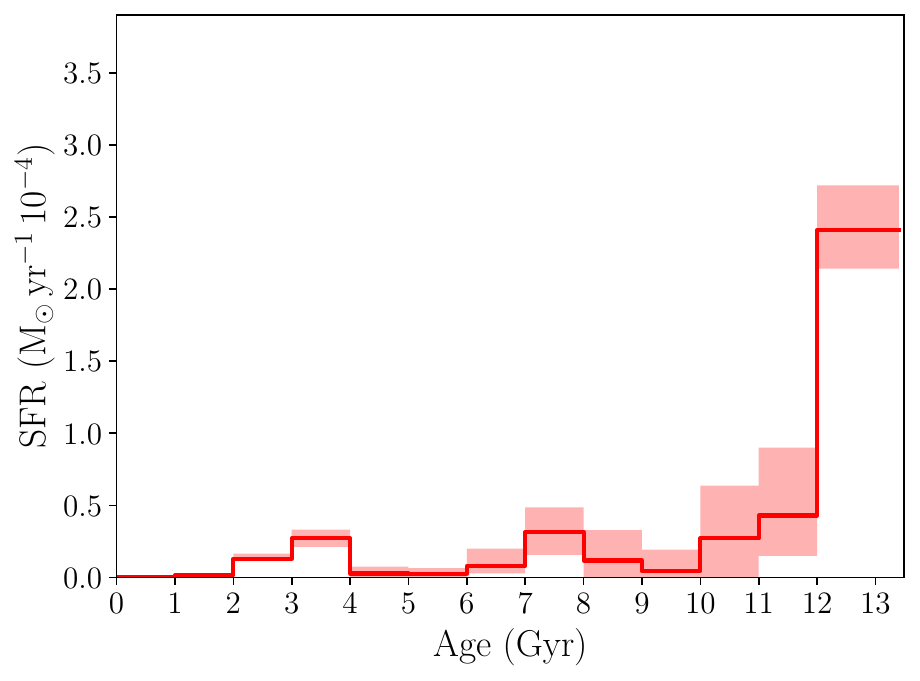}
    \caption{SFH of Terzan~5 associated to the SFERA best-fit solution. The red shaded areas are the uncertainties associated to the model computed from a bootstrap method, and they mark the $5\text{th}$ and $95\text{th}$ confidence level. To obtain this result, an average foreground extinction of $E(B-V)=2.5$ and distance modulus $\mu_0=13.75$ were adopted.}
    \label{fig: sfh_best}
\end{figure}

The result of the differential reddening correction and PM selection procedures discussed in Sects.~\ref{sec: ter5_sfh_reddening} and \ref{sec: field_contamination} is shown in the left panel of Figure~\ref{fig: cmd_drc}. By taking into account the completeness limits and to work with a suitable photometric sample, we run SFERA considering the CMD shown in the right panel of the figure, which includes only stars lying at distances larger than $15\arcsec$ from the center with optimal values of $\chi$ and $sharpness$ (i.e., stars not excluded by the $3\sigma$ clipping procedure, see Figure~\ref{fig: ter5_sfh_photo}). We also limited the analysis to stars brighter than $\rm I=22$, which corresponds to a completeness larger than 50\% in the considered area. Following this selection, the final sample of stars amounts to 8104.

\begin{figure}
\centering
    \hspace{0.1cm}
    \includegraphics[width=7.65cm]{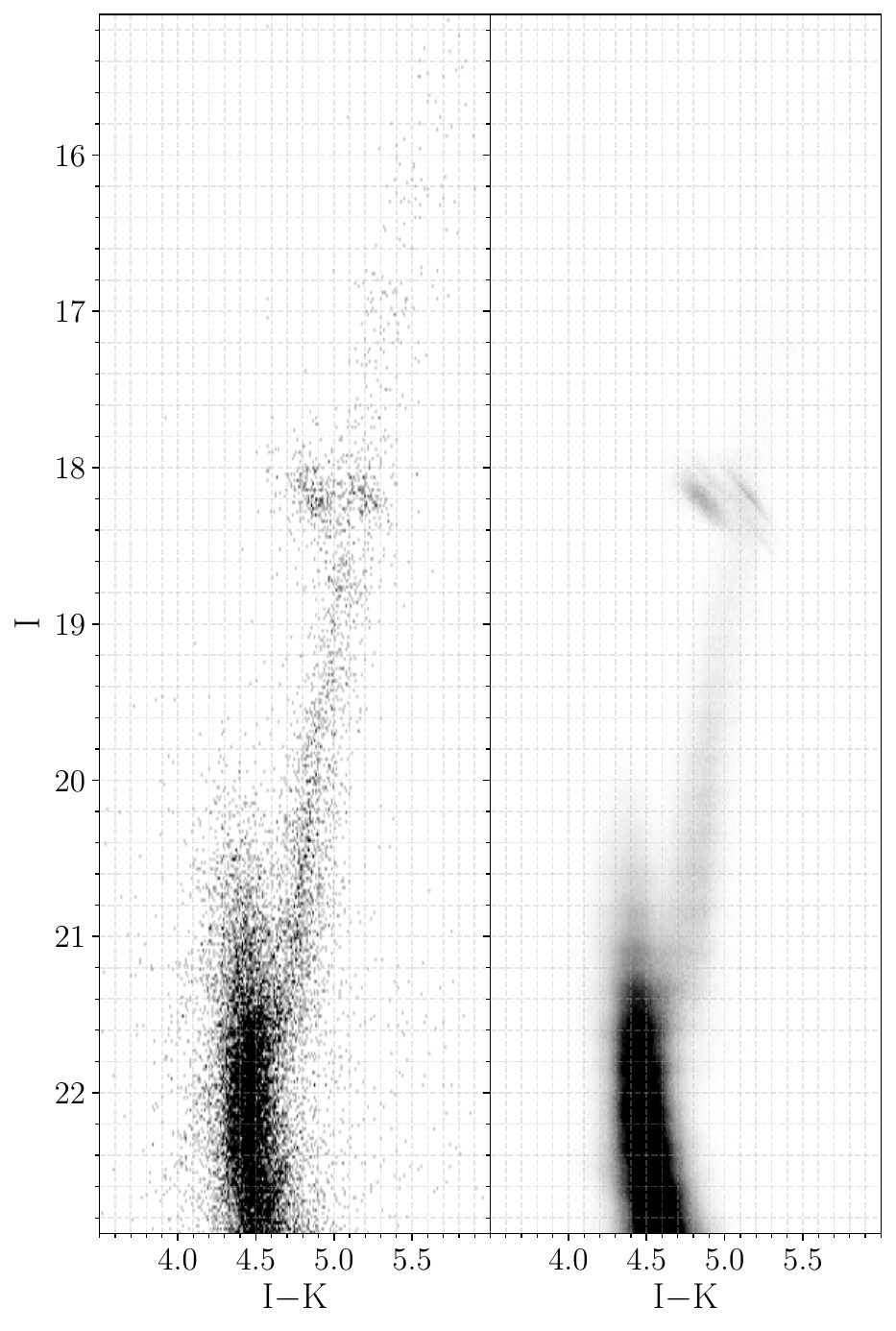}

    \includegraphics[width=8cm]{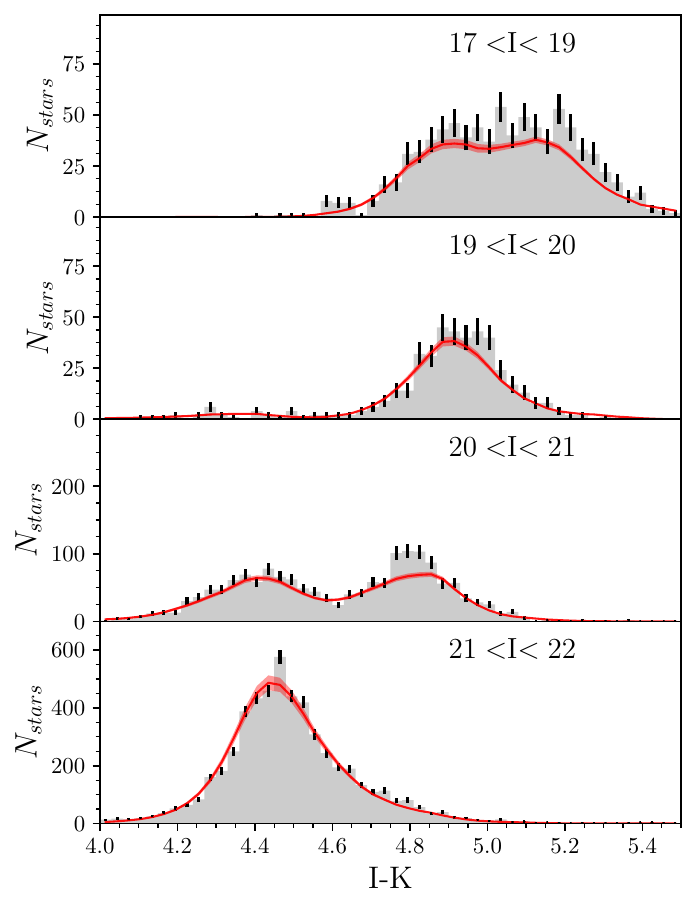}

    \caption{\emph{Upper panels:} Hess diagrams of the observed DRC (left) and best-fit synthetic (right) CMD. \emph{Lower panels:} Color functions of the CMDs in four different magnitude bins; the data and relative Poissonian uncertainty are displayed in gray, while the models are the superimposed red line.} 
\label{fig: hess_cfs_best}
\end{figure}

SFERA found a best-fit extinction corrected distance
modulus $\mu_0=13.75$ and average foreground reddening $E(B-V)=2.5$. The distance modulus agrees, within the uncertainties, with the value quoted in \cite{valenti+2007}, while the color excess is $\sim0.1\,$mag higher, being only marginally in agreement with the $E(B-V)$ calculated in the same work\footnote{In this respect, it is important to specify that, in the case of \cite{valenti+2007}, the mean color excess associated to Terzan~5 was determined from the pure NIR plane with no differential reddening corrections applied to the CMD.}. However, it is worth recalling the remarkable variation of $E(B-V)$ across the FOV of the observations (see Section~\ref{sec: ter5_sfh_reddening}, and also \citealt{massari+2012}).

The red line in Figure~\ref{fig: sfh_best} shows the best-fit SFH. The shaded areas represent the statistical uncertainties, corresponding to the 5th and 95th percentiles of the distribution of all the
synthetic CMDs produced by SFERA.
The SFH of Terzan~5 exhibits a main burst of star formation (SF) at $t\geqslant12\,$Gyr, that confirms the age of the old population inferred by the previous isochrone fitting analysis (\citealt{ferraro+2016}). Then, for younger ages, we can appreciate a prolonged star formation composed of several minor peaks, with the most relevant ones occurring around $8-7\,$Gyr, and between $4-2\,$Gyr ago.

\begin{figure}
    \centering
    \includegraphics[width=9cm]{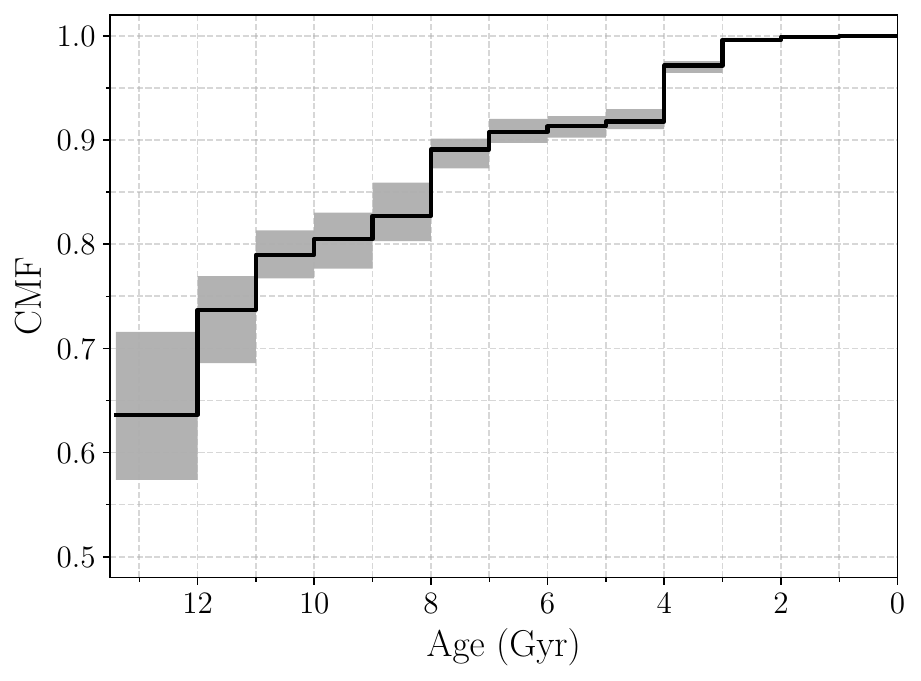}
    \caption{Cumulative stellar mass function for the recovered best-fit SFH.}
    \label{fig: cumulative_best}
\end{figure}

The upper plot of Figure~\ref{fig: hess_cfs_best} displays the observed data (left-hand panel) and the Hess diagram for the synthetic CMD, corresponding to the best-fit SFH (right-hand panel). From a visual inspection, the best-fit solution nicely reproduces both the extension and the color distribution of the blue plume, the location of the old MS-TO, and the complexity of the RC region. To compare the observed and synthetic CMDs in a more quantitative way, in the lower panel of Figure~\ref{fig: hess_cfs_best} we display the observed (gray histogram) and best-fit solution (red lines) color distributions in four different magnitude bins. Overall, the model color function does a fairly good job in reproducing both the width and the mean color distributions in all magnitude bins. 

The cumulative stellar mass function (CMF) associated to this solution is reported in Figure~\ref{fig: cumulative_best}. It shows that the first burst of SF builts $\sim65\%$ of the observed stellar mass. Afterward, the CMF of Terzan~5 gradually grows reaching the $\sim 90\%$ at $t<8\,$Gyr. The missing $10\%$ was then built-up during the latest SF episode (2-4 Gyr ago). The fraction of mass built during the young SF episode barely reconciles with the estimate from \cite{lanzoni+2010}, who concluded, based on the RC star counts, that the young population accounts for $\sim38\%$ of the total mass. However, it is important to note that, due to observational limitations, we excluded from the analysis the innermost $15\arcsec$ from the cluster center, where the metal-richer and younger population is preferentially concentrated (\citealt{ferraro+2009,lanzoni+2010}). Hence, the fraction of mass ascribable to the young stellar populations is inevitably underestimated in this analysis.

\begin{figure}[!t]
    \centering
    \includegraphics[width=9cm]{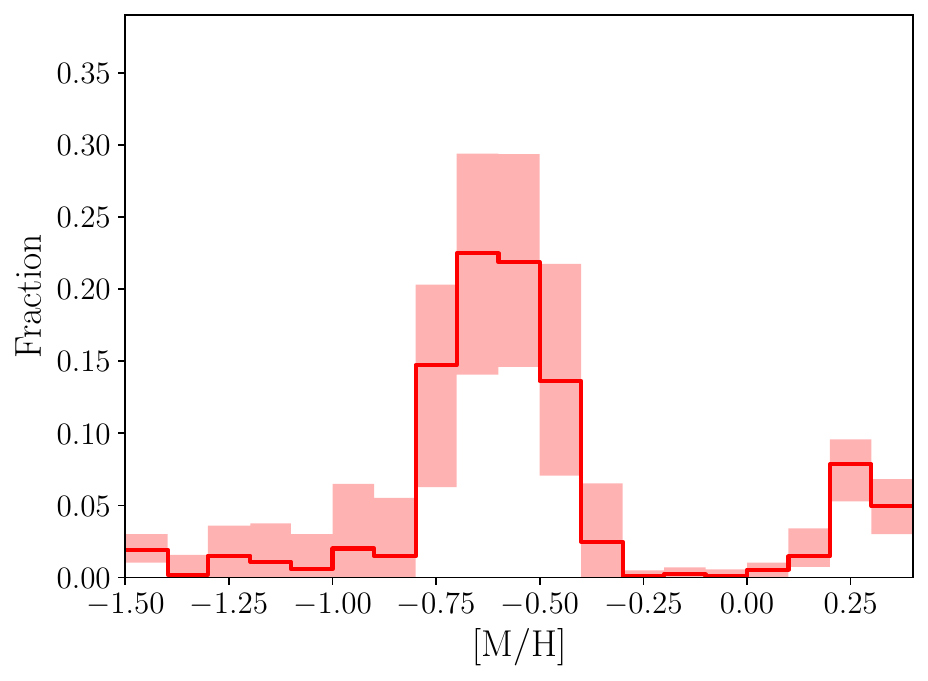}
    \caption{Metallicity distribution function as derived from the best-fit solution.}
    \label{fig: nz_best}
\end{figure}

Finally, in Figure~\ref{fig: nz_best} we show the metallicity distribution function (MDF) recovered from the best-fit solution. The distribution turns out to be clearly bimodal with two major components: a subsolar peak at $-0.7 < \rm [M/H] < -0.5\,$dex, 
and a supersolar one at $\rm [M/H] \sim +0.25\,$dex.
Qualitatively, the two metallicity peaks correspond to the the oldest and youngest star formation episodes, but a one-to-one association is not expected because intermediate bursts may introduce some spreads, instead of further isolated peaks.
The derived distribution appears to nicely reproduce the overall shape of the one obtained by \cite{origlia+2013,massari+2014} by using high-resolution spectroscopy of about 200 stars, including also the minor metal poor sub-component discussed in \cite{origlia+2013}\footnote{Small differences in the absolute value of the metallicity of the subsolar components are possibly due to a variety of factors, such as the adoption of solar-scaled isochrones, systematic offsets between isochrones and observations, and a possible metallicity-reddening degeneracy, which can be non-negligible in the extreme reddening conditions of Terzan~5.}.

\section{Summary and conclusions}
\label{sec: summary_conclusions}
The photometric and spectroscopic observations recently gathered for two massive and metal-rich stellar clusters in the Galactic bulge \-- Terzan~5 and Liller~1 \-- disclosed their intriguing peculiarity of hosting multiple stellar populations with remarkable metallicity ($\Delta \text{[Fe/H]}>0.5\,\text{dex}$) and age ($\Delta t>4\,\text{Gyr}$) spreads (\citealt{ferraro+2009,ferraro+2016,ferraro+2021,origlia+2011,origlia+2013,massari+2014,dalessandro+2022,crociati+2023}). At present, these are the only in-situ (see \citealt{massari+2019,callingham+2022}) stellar systems showing an old and subsolar population coexisting with a young and supersolar one, and their origin is still debated. 

The first hypothesis links these systems to one of the mechanisms that possibly contributed to the formation of the Milky Way's bulge (\citealt{ferraro+2009,ferraro+2021}). Specifically, massive and highly star-forming complexes, customarily dubbed clumps, are commonly observed in high-redshift, still bulgeless disk galaxies ($z\ge1.5$; see e.g., \citealt{guo+2015}). An increasing number of observational (\citealt{guo+2018,ambachew+2022}) and theoretical (e.g., \citealt{elmegreen+2008,bournaud2016,garver+2023}) studies are suggesting that these disk structures can migrate to the center of the host galaxy and contribute to the assembly of its bulge. A small fraction of them is also predicted to survive this coalescence process and be nowadays observable as cluster-like stellar systems (\citealt{dekel+2023}). Therefore, Terzan~5 and Liller~1 can represent the present-day remnants of two such massive clumps. Given their high progenitor mass, these systems may have undergone a self-enrichment process.
An alternative hypothesis, firstly advanced by \citet[][see also \citealp{bastian+2022}]{mckenzie+2018} through numerical simulations, involves the gravitational capture of field GMCs by old massive GCs. Under highly fine-tuned conditions (regarding the orbits, relative velocities, masses and dimensions of the systems, the number of GMCs supposed to be present in the bulge at the time of the merging, etc.), a strongly bound GC-GMC collision would fuel subsequent a burst of star formation. However, this is an intrinsically rare event, which very unlikely can happen more than once in a cluster life.

An  efficient way to disentangle between these two formation scenarios is the study of the SFH of these systems, to verify whether they experienced just one or more star formation bursts, or even a prolonged activity. Indeed, the reconstructed SFH of Liller~1 (\citealt{dalessandro+2022}) shows evidence of multiple and long-lasting star formation bursts. Motivated by those results, here we conducted an analogous investigation on Terzan~5. To retrieve its SFH, we employed the code SFERA comparing the observed CMD of Terzan~5 with a set of synthetic ones through the CMD fitting technique.
The observed CMD used in this comparison has taken advantage of the powerful combination of deep optical (HST) and IR Adaptive Optics (VLT/MAD) observations. Indeed, the hybrid (I, I$-$K) CMD turns out to be the ideal tool for obtaining a clear view of the stellar evolutionary sequences, especially in regions of strong reddening as the Galactic bulge. In addition, such a data set allowed the accurate determination of relative PMs for the removal of Galactic field interlopers, and the construction of a high spatial resolution (below $1.5\arcsec$) reddening map, used to correct for the effects of differential extinction.

The best-fit solution shows a main and narrow burst of SF around $13\,$Gyr ago, combined with a decreasing and prolonged SF activity followed by another significant SF episode starting $4\,$Gyr ago and extending for $\sim2\,$Gyr. A possible additional burst occurred $\sim7-8\,\rm Gyr$ ago. The overall time extension of the main bursts is remarkable, and it would point to a constant supply of gas to fuel a SF activity with a duration of more than $1\,$Gyr. Beside the details, the cumulative mass function derived from this study clearly shows a gradual build-up of the stellar mass in this system, which suggests a continuous, low rate SF activity characterized by a few main bursts. 

The overall structure of the reconstructed SFH of Terzan~5 appears astonishingly similar to that derived for Liller~1 \citep{dalessandro+2022}, showing an uninterrupted star formation activity with at least three relevant star formation events in both systems. The first, major episode appears to be quite prolonged in time (1-2 Gyr), at odds with what observed for genuine GCs. This is followed by two additional bursts emerging from an underlying continuous star formation activity with very low intensity. The most recent bust occurs at different look-back times in the two systems (1-2 Gyr ago in Liller 1, and 4-5 Gyr ago in Terzan 5). It has been possibly triggered by violent interactions with other Bulge substructures (as the bar) and thus likely depends on the details of the orbits of each stellar system. Intriguingly, instead, the intermediate-age burst occurred at the same cosmic age in both the system (8-9 Gyr ago). This curious synchrony of the two events needs to be further investigated since it might trace a major event in the past of the Galactic Bulge. In both cases, the reconstructed SFH is hardly reconcilable with the predicted short timescales ($<10\,$Myr, \citealt{mckenzie+2018}) needed to a GMC to form stars after a bound collision with a GC. The results presented in this paper thus further support a scenario where the complexity  of Terzan~5 and Liller~1 is due to self-enrichment processes in massive structures that contributed to form the bulge of our Galaxy.

\begin{acknowledgements}
This work is part of the project Cosmic-Lab ("Globular Clusters as Cosmic Laboratories") at the Physics and Astronomy Department "A. Righi" of the Bologna University (\url{http://www.cosmic-lab.eu/Cosmic-Lab/Home.html}). C.C. acknowledges the ESO Studentship Programme for funding the period spent at the ESO Headquarters in Garching bei München (Germany) where part of this work was carried out. E.V. acknowledges the Excellence Cluster ORIGINS funded by the Deutsche Forschungsgemeinschaft
(DFG, German Research Foundation) under Germany's Excellence Strategy \-EXC\-2094\-39078331.
\end{acknowledgements}

\bibliography{aanda}{}
\bibliographystyle{aa}

\end{document}